\documentclass[epj]{svjour}
\usepackage{graphics}
\usepackage{times}
\usepackage{helvet}
\def\ga{\mathrel{\mathchoice {\vcenter{\offinterlineskip\halign{\hfil
$\displaystyle##$\hfil\cr>\cr\sim\cr}}}
{\vcenter{\offinterlineskip\halign{\hfil$\textstyle##$\hfil\cr>\cr\sim\cr}}}
{\vcenter{\offinterlineskip\halign{\hfil$\scriptstyle##$\hfil\cr>\cr\sim\cr}}}
{\vcenter{\offinterlineskip\halign{\hfil$\scriptscriptstyle##$\hfil\cr>\cr
\sim\cr}}}}}
\begin{document}
\title{Transverse mass distributions of neutral pions 
from $^{208}$Pb-induced reactions at 158$\cdot A$~GeV}
\subtitle{WA98 Collaboration}
\author{
M.M.~Aggarwal\inst{4}
\and A.L.S.~Angelis\inst{7} 
\and V.~Antonenko\inst{13}
\and V.~Arefiev\inst{6}
\and V.~Astakhov\inst{6}
\and V.~Avdeitchikov\inst{6}
\and T.C.~Awes\inst{16}
\and P.V.K.S.~Baba\inst{10}
\and S.K.~Badyal\inst{10}
% ...
}
\institute{
Institute of Physics, 751-005  Bhubaneswar, India
\and Variable Energy Cyclotron Centre,  Calcutta 700 064, India
\and MIT Cambridge, MA 02139, USA 
\and University of Panjab, Chandigarh 160014, India
\and Gesellschaft f{\"u}r Schwerionenforschung (GSI), D-64220 Darmstadt, Germany 
\and Joint Institute for Nuclear Research, RU-141980 Dubna, Russia
\and University of Geneva, CH-1211 Geneva 4,Switzerland
\and KVI, University of Groningen, NL-9747 AA Groningen, The Netherlands 
\and University of Rajasthan, Jaipur 302004, Rajasthan, India
\and University of Jammu, Jammu 180001, India
\and University of Tennessee, Knoxville, Tennessee 37966, USA
\and Lund University, SE-221 00 Lund, Sweden 
\and RRC ``Kurchatov Institute'', RU-123182 Moscow, Russia
\and University of M{\"u}nster, D-48149 M{\"u}nster, Germany
\and SUBATECH, Ecole des Mines, Nantes, France
\and Oak Ridge National Laboratory, Oak Ridge, Tennessee 37831-6372, USA
% ...
}
\abstract{
Results on transverse mass spectra of neutral 
pions measured at central rapidity 
are presented for impact parameter selected 
158$\cdot A$~GeV Pb\,+\,Pb, and Pb\,+\,Nb collisions.  The 
distributions cover the range $0.5 \, \mathrm{GeV}/c^{2} \le 
m_{T} - m_{0} \le 4 \, \mathrm{GeV}/c^{2}$. 
The change of the spectral shape and the multiplicity 
with centrality is studied in 
detail. In going from p+p to semi-peripheral Pb+Pb collisions there is 
a nuclear enhancement increasing with transverse mass similar to the 
well known Cronin effect, while for very central collisions this 
enhancement appears to be weaker than expected.
} 
\maketitle
\begin{figure*}[bt]
   \centerline{\includegraphics{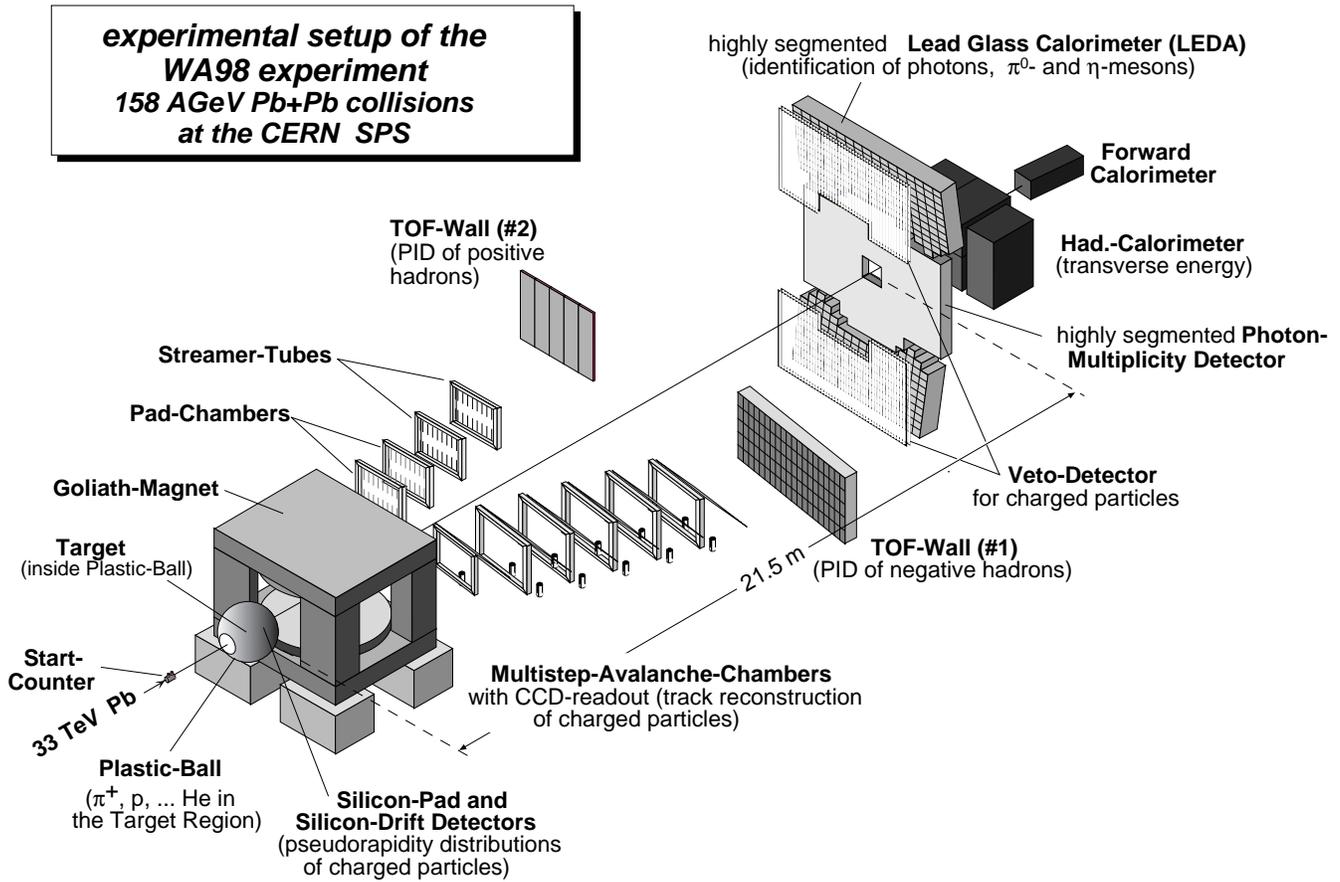}}
   \caption{The WA98 experimental setup.}
   \protect\label{fig:wa98}
\end{figure*}
\section{Introduction}
Heavy ion reactions at ultrarelativistic energies 
provide information on nuclear matter at high energy density
(for reviews see e.g. \cite{qm96,qm97,qm99}). Hadron 
production is generally considered to be sensitive to the late
freeze-out stage
of the collision, when hadrons 
decouple from one another. Already from the experimentally 
determined shape of the 
transverse mass spectra it is evident that heavy ion reactions are 
not merely a superposition of nucleon-nucleon collisions 
\cite{Jones:1996xc,na44:coll,wa80:pi0:98}.

The broadening of the transverse mass spectra in p+A collisions 
compared to p+p (Cronin-effect \cite{cronin}) has been attributed to 
\emph{initial state} multiple scattering of partons \cite{krzywicki79,LEV83}.
However, models which attempt to describe nucleus-nucleus reactions, like 
the Monte-Carlo programs VENUS \cite{WERNER93a} or RQMD \cite{sorge89} 
rely on the assumption that \emph{final state} rescattering plays an 
important role in determination of the momentum distributions of the 
hadrons. If final state 
rescattering is very strong, the 
notion of local kinetic equilibrium may be valid, which 
is the basic assumption for a
hydrodynamical description (see e.g. \cite{LEE89,schnedermann93a}).

\begin{sloppypar}
It remains unclear, whether 
equilibration is attained.  If it occurs, it will be important to
isolate the 
contributions of thermal, pre-equilibrium, and initial state
processes to the hadron yield. Since the 
latter are expected to dominate at large transverse momenta, it is 
hoped that systematic studies of hadron spectra over a large  
range in momentum might allow to disentangle these contributions. 
In the analysis of central reactions of Pb~+~Pb at 158$\cdot A$~GeV it is 
seen that both predictions of perturbative QCD 
\cite{wa98:pi0:98,wang:1998:qcd} and hydrodynamical parameterizations 
\cite{wa98:hydro:99} can describe the measured neutral pion spectra 
reasonably
well. It is particularly surprising to observe that, on the one hand, 
a pQCD calculation gives a reasonable description also at relatively low momenta, 
while on the other hand, a hydrodynamical parameterization can
provide a good description, even at very high momenta.
\end{sloppypar}

The understanding of the relative contributions of the various soft and 
hard processes in particle production is especially important in view 
of the recent interest in the energy loss of partons in dense matter 
\cite{jetq1,jetq2}, 
generally referred to as \emph{jet quenching}, as a possible probe 
for the quark-gluon-plasma. One of the suggested experimental hints 
of jet quenching is the suppression of particle production at high 
transverse momenta \cite{jetq3}. 
In order to confirm such an interpretation, it is 
important to study other possible 
nuclear modifications of particle production in detail.
More information in this respect 
may be gathered from the variation of the particle spectra for 
different reaction systems or different centralities. First attempts 
in this respect have been discussed in \cite{wa98:pi0:98}, where the 
neutral pion average $p_{T}$ as a function of the centrality was shown to 
rise from 
p+p collisions to peripheral Pb+Pb collisions and to saturate for 
medium central and central Pb+Pb collisions. It was also seen that 
the neutral pion yield for reactions with more than about 30 
participating nucleons $N_{part}$ exhibits a scaling as
$N_{part}^{\alpha}$ with a power $\alpha \approx = 1.1$ which is 
approximately independent of $p_{T}$. 

In the present paper we will present a detailed study of neutral 
pion transverse mass spectra in the range 
$0.5 \mathrm{GeV}/c^{2} \le m_T - m_0 \le 4.0 \mathrm{GeV}/c^{2}$ and 
$2.3 \le y \le 3.0$ for collisions of Pb+Pb and Pb+Nb at 158$\cdot A$~GeV 
for different centralities.

\section{Experiment}

The CERN experiment WA98 \cite{misc:wa98:proposal:91} is a 
general-purpose apparatus which consists
of large acceptance photon and hadron spectrometers together with several 
other large acceptance devices which allow to measure various global
variables on an event-by-event basis. The experiment took data
with the 158$\cdot A$~GeV $^{208}$Pb beams from the SPS in 1994, 1995, and
1996. The data presented here were taken during the 1995 and 1996 
beamtimes at the
CERN SPS. 
The layout of the WA98 experiment as it existed during the final WA98 
run period in 1996 is shown in Fig.~\ref{fig:wa98}. 

\begin{sloppypar}
Neutral pions are reconstructed on a statistical basis 
from their two-photon decay, using 
photons measured with the LEDA
spectrometer in the pseudorapidity interval $2.3 < \eta < 3.0$.
This detector is located 21.5~m from the target and consists of 10,080 
modules. Each module is a $4 \times 4 \times 40$~cm$^3$ 
(14.3 radiation lengths) TF1 lead-glass
block read out by an FEU-84 photomultiplier. The high voltage is generated
on-base with custom developed  \cite{ref:ne95} Cockcroft-Walton 
voltage-multiplier type bases which are
individually controlled by a VME processor.
The photomultiplier signals are digitized with a custom-built ADC system
\cite{ref:wi94}.
Twenty-four lead-glass modules are epoxied together in an array
6 modules wide by 4 modules high to form a super-module. Each
super-module has its own calibration and gain monitoring system based on
a set of 3 LEDs and a PIN-photodiode mounted inside a sealed reflecting 
front cover dome \cite{ref:pe96}. 
The photon spectrometer is separated
into two nearly symmetric halves above and below the beam plane
in the two regions of reduced charged-particle density which result from the 
sweeping action of the GOLIATH magnet.
More details about the experimental setup and the photon spectrometer can 
be found in \cite{wa98:photons}.
\end{sloppypar}

\begin{figure}[bt]
        \includegraphics{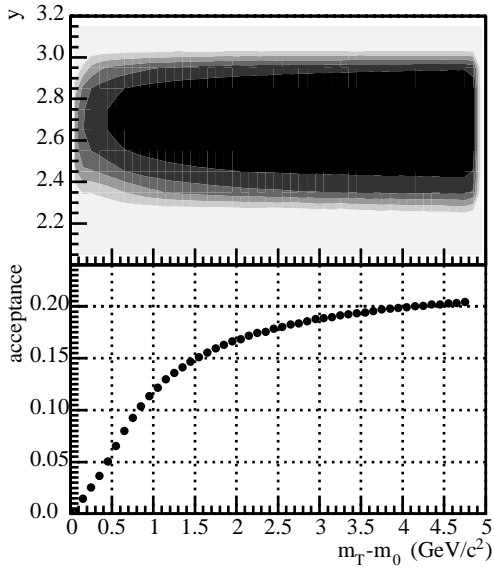}
        \caption{Top: Geometrical acceptance for neutral pions as a function of 
        $p_{T}$ and $y$. Bottom: Rapidity-integrated acceptance in 
        $2.0 < y < 
        3.2$ as a function of $m_{T} - m_{0}$.}
        \protect\label{fig:acceptance}
\end{figure}

The acceptance of the photon spectrometer for $\pi^{0}$  
detection in rapidity and transverse mass is
shown in Fig.~\ref{fig:acceptance}. The
acceptance covers the region $2.3 < y < 3.0$,  near mid-rapidity 
($y_{cm}=2.9$).

\begin{sloppypar}
The minimum bias trigger requires a valid signal of the beam 
counters
and a minimum amount of transverse energy $E_T \ga 
5$\,GeV, detected by the Mid-Rapidity Calorimeter, MIRAC~\cite{mirac}, 
which is located 24~m downstream of the target. 
MIRAC consists of a hadronic and an electromagnetic 
section and covers the pseudorapidity interval 
$3.5 < \eta < 5.5$.
Data have been taken with the 158$\cdot A$~GeV 
lead beam on targets of Pb (495 and 239 mg/cm$^2$) and Nb
(218~mg/cm$^2$).  
For the present analysis $9.7$ million Pb+Pb and 
$0.23$ million Pb+Nb minimum bias events were accumulated. 
 The minimum bias 
cross sections have been calculated from the number of beam triggers and 
minimum bias triggers and the target thicknesses. The yields have been 
corrected for small non-target background contributions 
(typically a few percent) to obtain $\sigma_{\mathrm mb} 
\approx 6300$\,mb and 4400\,mb for Pb\,+\,Pb and Pb\,+\,Nb reactions, 
respectively. 
These absolute cross sections are estimated to have an overall 
systematic error of less than 10\%.
\end{sloppypar}

\section{Pion Reconstruction and Efficiency}
\label{sec:rec}

The method for pion reconstruction is similar to that used by 
the WA80 collaboration as discussed in \cite{wa80:pi0:98}. It 
is also discussed at  length in \cite{wa98:photons,phd:blume} 
so we will only briefly sketch the basics of the method.

The showers used for the extraction of the neutral pion yield via the 
$\gamma\gamma$ decay branch can be selected with different criteria. 
It has been demonstrated that the extracted yield does not depend 
significantly on the choice of these photon identification cuts 
\cite{wa98:photons}. For the present analysis,  
showers with a small lateral dispersion \cite{berger92} have
been selected as photon candidates.

\begin{figure}[bt]
        \includegraphics{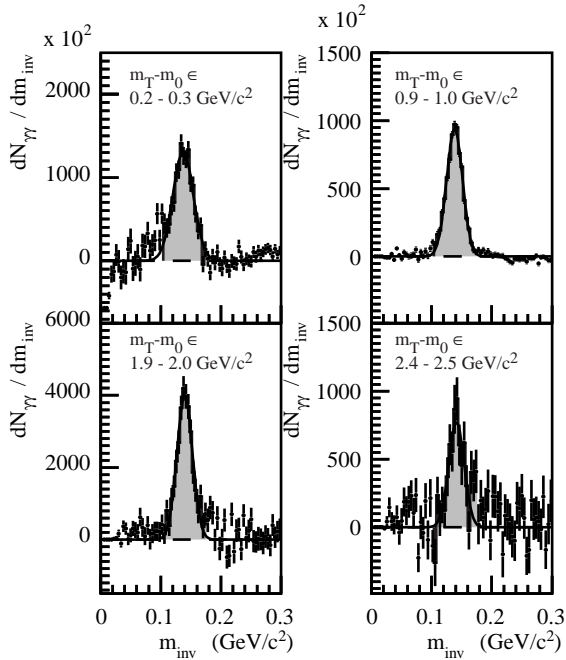}
        \caption{Invariant mass distributions of photon pairs for minimum 
        bias Pb~+~Pb collisions in different intervals of $m_{T} - m_{0}$. 
        The combinatorial background has been calculated using 
        mixed-event distributions and subtracted.}
        \protect\label{fig:pipeak}
\end{figure}

\begin{sloppypar}
Hits in LEDA are combined pairwise to provide 
distributions of pair mass vs.\ $m_{T}-m_{0}$ (where 
$m_{T} = \sqrt{p_T^2+m_0^2}$ is the transverse mass
and $m_0$ is the $\pi^0$ rest mass) for all 
possible combinations.  These distributions are obtained both for 
real events, $R(m_{\mathrm{inv}},m_{T})$, and for so-called mixed 
events, $M(m_{\mathrm{inv}},m_{T})$, where a hit from one event 
is combined with a hit from another event with similar 
multiplicity.  $M(m_{\mathrm{inv}},m_{T})$ provides a good 
description of the combinatorial background and is subtracted 
from $R(m_{\mathrm{inv}},m_{T})$ to obtain the mass distribution 
of neutral pions (see Fig.~\ref{fig:pipeak}).
\end{sloppypar}

The large multiplicities, especially in central reactions of 
Pb~+~Pb, lead to a considerable probability that showers in the 
detector overlap and influence each other. Particles may be lost for 
reconstruction. In other cases the particle might be measured with an 
incorrect energy.
These effects lead to a detector efficiency for pion 
reconstruction which depends on particle density. To study this 
detection efficiency, the GEANT \cite{geant} simulation package  
has been used to create artificial signals for the lead glass 
modules corresponding to neutral pions incident on the detector. 
The effects of detector noise and the digitization of the photomultiplier 
signals are also implemented.  
To obtain 
the $\pi^{0}$ reconstruction efficiency, 
these simulated photon shower pairs have been 
superimposed onto the measured events.  
The simulation provides the 
means to extract the probability that a pion at a given input 
transverse mass $m_{T}^{(0)}$ will be measured with 
$m_{T}^{(1)}$.  These probabilities, extracted as a function of 
$m_{T}^{(0)}$, can be understood as a matrix which transforms the real 
physical distributions into the measured ones. 
These correction matrices are used to correct the measured 
distributions. A detailed description of the efficiency correction can 
be found in \cite{wa98:photons}.

The distributions have then been corrected for contributions from 
reactions of Pb projectiles with material other than the target (e.g. 
residual gas, exit windows, etc.). The corresponding corrections were 
obtained by measurements performed without target. The corrections 
are negligible for medium-central and central reactions.

The systematic errors on the measured transverse mass spectra are 
dominated by the following contributions:
\begin{itemize}
    \begin{sloppypar}
        \item An uncertainty in the absolute calibration of the momentum 
        scale of 1\%. This may be translated into an uncertainty in the 
        yields 
        of a few \% at low $m_{T}$ rising to $\approx 13 \%$ at 
        \mbox{$m_{T} - m_{0} = 3.5 \, \mathrm{GeV}/c^{2}$}, 
	which is independent of centrality 
        or reaction system.
	\end{sloppypar}
        
        \item Uncertainties in the $\pi^{0}$ extraction which include the 
        error in the determination of the invariant mass peak content and 
        the propagation of the energy resolution through the acceptance and 
        efficiency corrections. This is largest in central 
        collisions. It leads to an error below 10\% for 
        $m_{T} - m_{0} \geq 0.5 \, \mathrm{GeV}/c^{2}$ for most samples. 
        
        \item Uncertainties in the correction for contributions from no-target 
        reactions, which are only relevant for peripheral reactions. 
        This leads to an error below 10\% for 
        $m_{T} - m_{0} \geq 0.5 \, \mathrm{GeV}/c^{2}$.
\end{itemize}
The total centrality dependent systematic error is below 10\% for 
$m_{T} - m_{0} \geq 0.5 \, \mathrm{GeV}/c^{2}$ for most centrality 
classes and below 20\% for the most central sample.

\begin{figure*}[tb]
   \centerline{\includegraphics{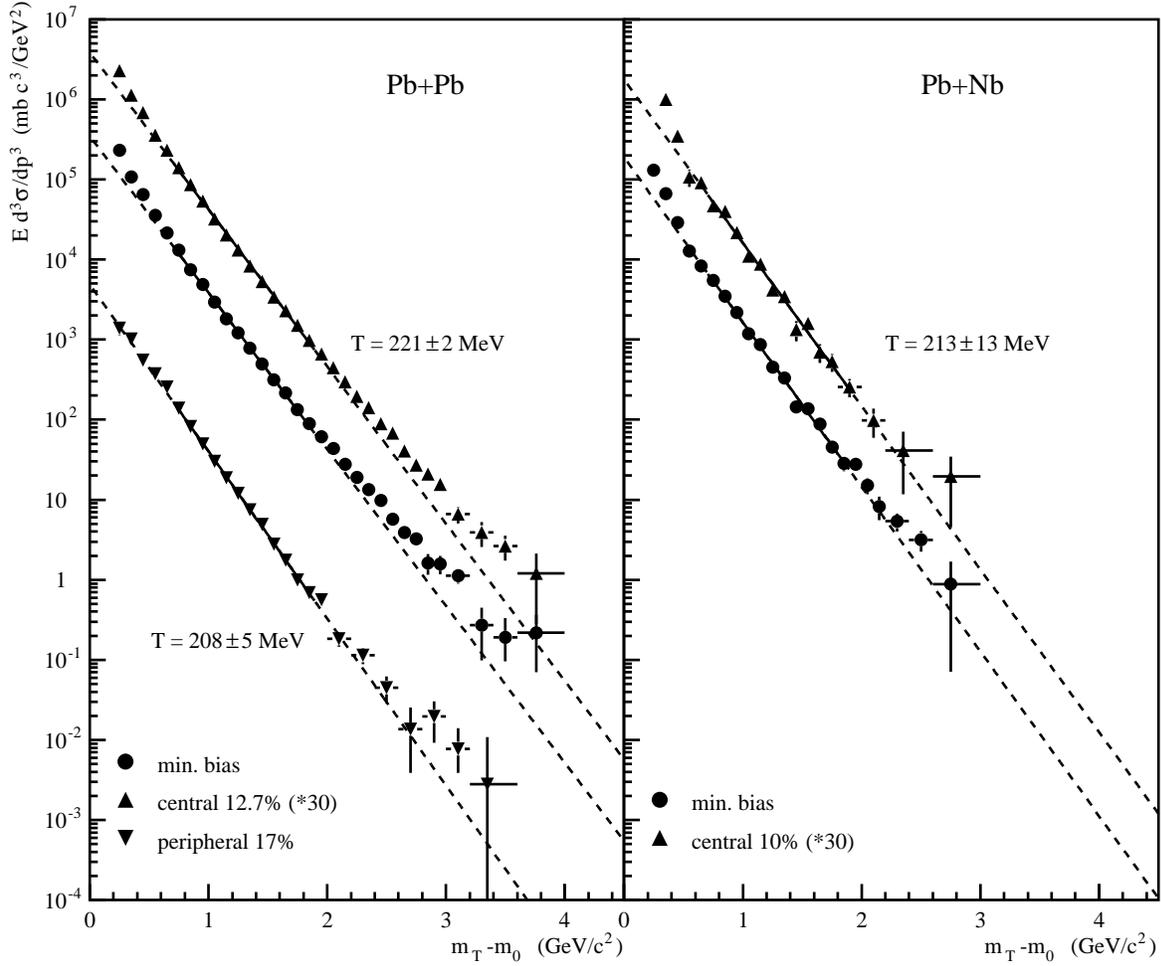}}
   \caption{Invariant cross sections of neutral pions 
        for Pb~+~Pb (left) and Pb~+~Nb collisions (right) of different centralities 
        as a function of $m_{T} - m_{0}$.}
   \protect\label{fig:spectall}
\end{figure*}

\section{Results}

\subsection{Neutral Pion Spectra}

Neutral pion spectra for Pb~+~Pb and Pb~+~Nb collisions for minimum 
bias and selected centralities
are presented in Fig.~\ref{fig:spectall}. On the left hand side results 
for Pb~+~Pb minimum bias as well as for the 12.7\% most 
central and the 17\% most peripheral event selections are shown. 
Included also are results of exponential fits:
\begin{equation}
        f(m_{T}-m_{0}) = C \cdot \exp \left(-\frac{m_{T}-m_{0}}{T}\right)
        \label{eq:exp}
\end{equation}
fitted over the range 
$0.7 \, \mathrm{GeV}/c^{2} \leq m_{T} - m_{0} \leq 1.9 \, \mathrm{GeV}/c^{2}$. 
As noted previously \cite{wa80:pi0:98,wa98:pi0:98}, the data clearly 
deviate from the exponential shape when considered over the full range of 
transverse masses. The fit in the limited transverse mass range can, 
however, still be used to extract slope parameters $T$. For the 12.7\% most 
central collisions one finds $T = 221 \pm 2 \, \mathrm{MeV}$ which is 
significantly larger than the value of 
$T = 208 \pm 5 \, \mathrm{MeV}$ for peripheral collisions.
Spectra for reactions of Pb~+~Nb (minimum bias and 10\% central) are 
shown on the right hand side. The spectral shapes are very similar
to those for  Pb~+~Pb collisions. 
For central collisions a slope parameter of $T = 213 \pm 13 \, 
\mathrm{MeV}$ is extracted which is intermediate compared to that of
peripheral and 
central Pb~+~Pb collisions. Since the statistics 
for the Pb~+~Nb dataset is limited, we will concentrate on the Pb~+~Pb 
reactions in the rest of the paper. 

\begin{figure}[htb]
        \centerline{\includegraphics{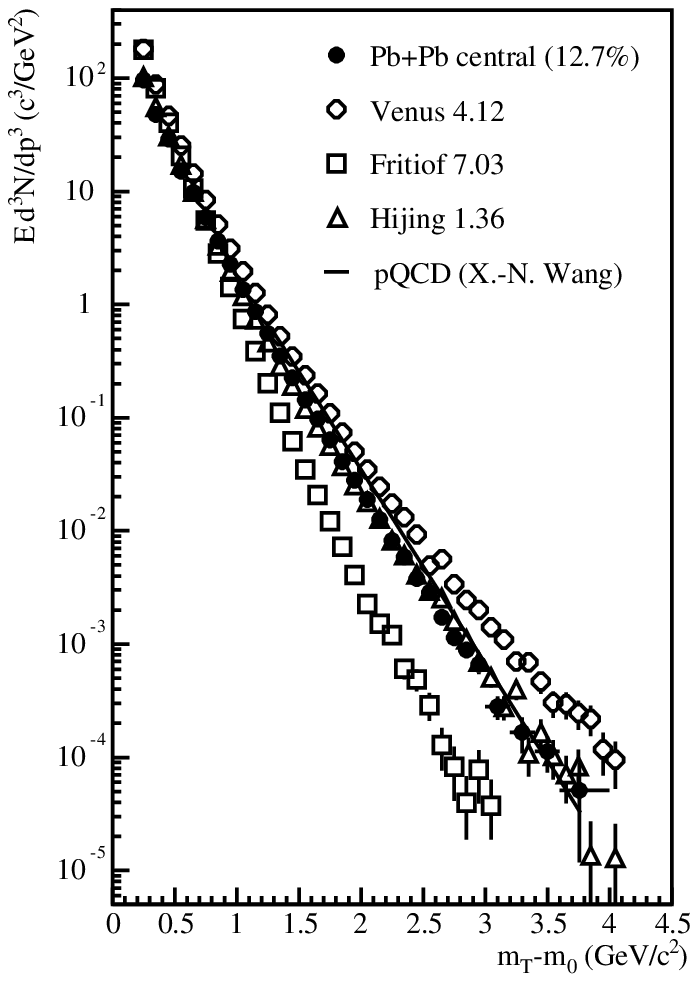}}
        \caption{Invariant multiplicities of neutral pions 
        for central Pb~+~Pb collisions as a 
        function of $m_{T} - m_{0}$ compared to predictions of Monte Carlo 
        event generators.        \protect\label{fig:spectmodel}}

        \centerline{\includegraphics{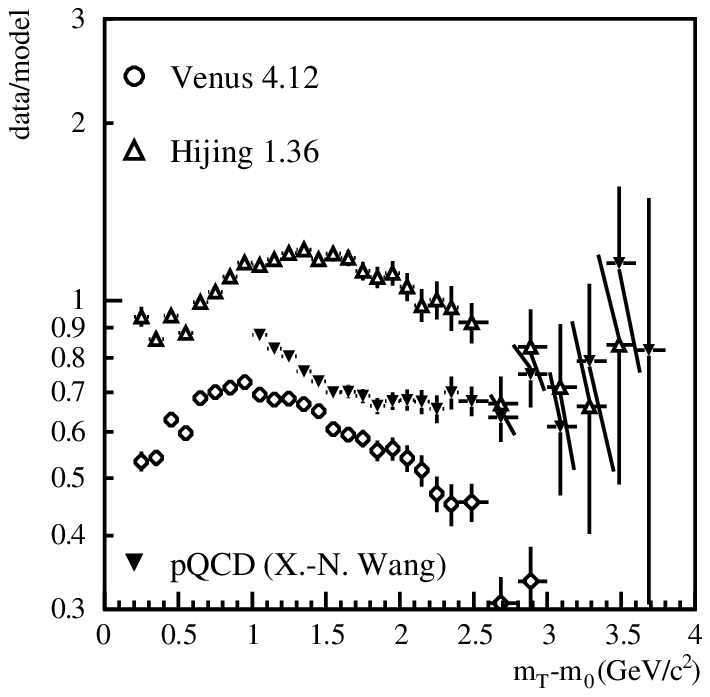}}
        \caption{Ratios of measured invariant multiplicities of 
        neutral pions to those from Monte Carlo 
        event generators
        for central Pb~+~Pb collisions as a 
        function of $m_{T} - m_{0}$.\protect\label{fig:ratiomodel}}
        
\end{figure}

In Fig.~\ref{fig:spectmodel} the multiplicity distribution for 
central collisions is compared to predictions of the event generators
FRITIOF 7.03 \cite{ANDERSSON93}, VENUS 4.12 \cite{WERNER93a}, and 
HIJING 1.36 \cite{wang:hijing}. Clearly FRITIOF 
does not describe the data at all, while VENUS and HIJING yield a 
more reasonable description. The prediction of the pQCD calculation 
from \cite{wang:1998:qcd} is included as a solid line, which also 
shows a reasonable agreement with the data. 
The degree of agreement of the models can be better 
seen in Fig.~\ref{fig:ratiomodel} where the ratio of the 
experimental data to the generator results is shown. 
FRITIOF is not included as the discrepancies are already evident from 
Fig.~\ref{fig:spectmodel}. VENUS overpredicts the pion production at 
high $m_{T}$ by about a factor of two. HIJING shows the best 
agreement at large momenta but underpredicts the data slightly at 
intermediate $m_{T}$. Also the Monte Carlo models lead to a 
stronger concave curvature than the data. The pQCD calculation 
overpredicts the data by $\approx 30 \% $ in the range 
$1.5 \, \mathrm{GeV}/c \le m_{T}-m_{0} \le 3.0 \, \mathrm{GeV}/c$.

\begin{figure}[bt]
        \centerline{\includegraphics{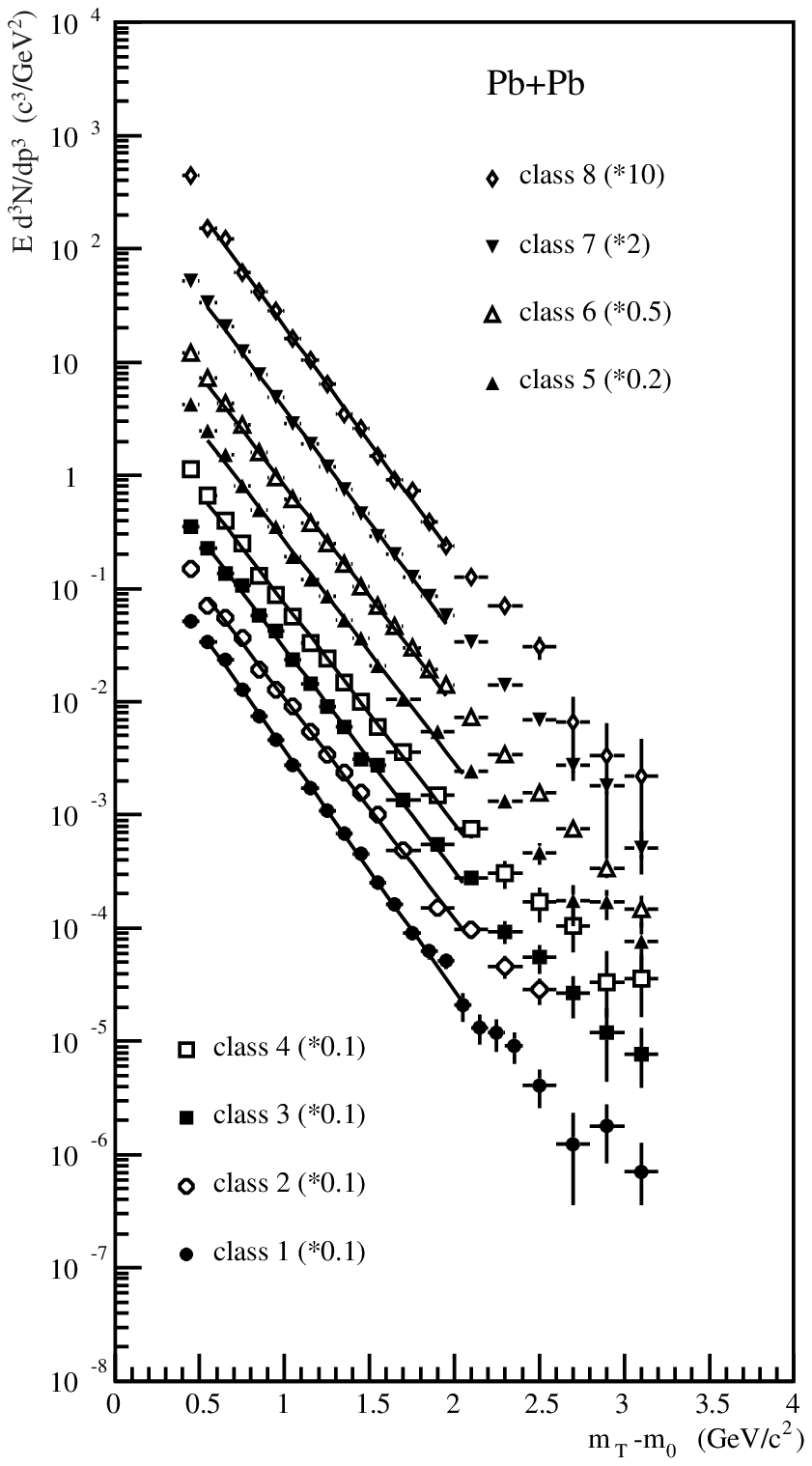}}
        \caption{Invariant multiplicities of neutral pions 
        for Pb~+~Pb collisions of different centralities as a 
        function of $m_{T} - m_{0}$. The solid lines show exponential 
        fits to the spectra.}
        \protect\label{fig:spectra1}
\end{figure}

The momentum spectra for central collisions have been published 
already in \cite{wa98:pi0:98}. As discussed in the corresponding erratum, 
cross section estimates had turned out to be incorrect. 
In the 
course of the present analysis it was realized that for the presentation
of the results in figure 1 of Ref. \cite{wa98:pi0:98}, the $\pi^{0}$ 
multiplicities were
incorrectly normalized. All other results and conclusions of
Ref. \cite{wa98:pi0:98} are unchanged. Unfortunately, the same 
improperly  normalized 
distributions were analyzed in Ref. \cite{wang:1998:qcd}. The 
correct normalization
reduces the $\pi^{0}$ mulitiplicities by $27 \% $ with the result 
that the pQCD calculations presented there would overpredict
the measured WA98 $\pi^{0}$ result, as shown in Fig.~\ref{fig:ratiomodel}.
Still  
different parameters for the $p_{T}$-broadening in these calculations 
might possibly enable the model to describe the central 
distributions better. More stringent tests of such a model can be 
performed when looking at the detailed centrality dependence of the 
pion production which has been done in the following.

The minimum bias sample has been subdivided into eight centrality 
samples summarized in Table~\ref{tab:classes}. Measurements were 
performed 
with and without magnetic field, which does not affect the neutral 
pion distributions, but alters the transverse energy used to determine 
the centrality. The corresponding cuts have been adjusted so that 
always the same fractions of the minimum bias cross section were 
selected. The data samples with and without magnetic field agree well 
with each other and have been combined in the present analysis. 
Table~\ref{tab:classes} shows the transverse energy cuts for 
one particular data set as an example.

\begin{figure}[bt]
        \centerline{\includegraphics{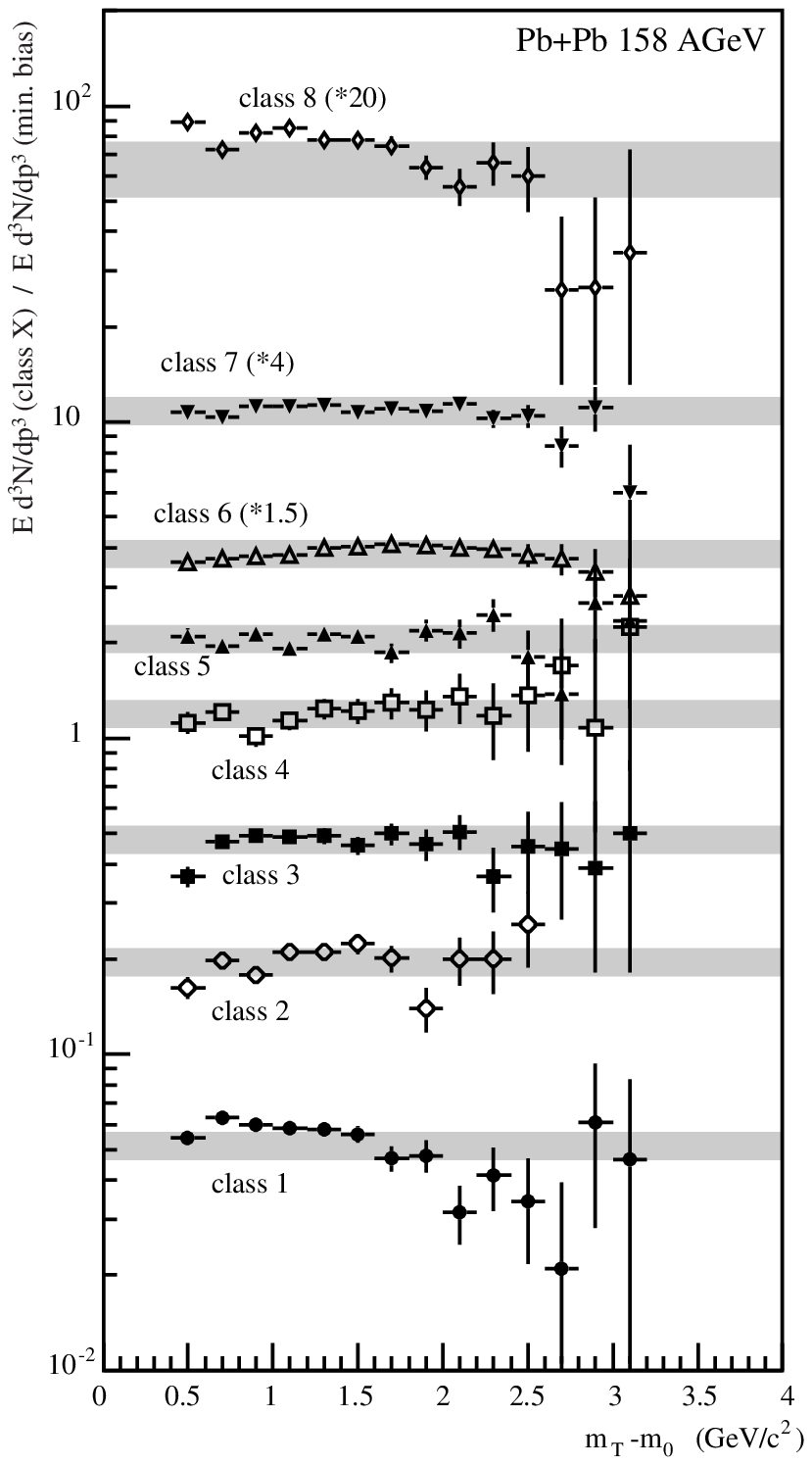}}
        \caption{Ratios of invariant multiplicities of neutral pions 
        for Pb~+~Pb collisions of different centralities to minimum bias 
        distributions as a 
        function of $m_{T} - m_{0}$, with class 1 being the most 
        peripheral and class 8 the most central sample. 
	The grey bands show the estimate 
        of the centrality dependent systematic error.}
        \protect\label{fig:ratioeff}
\end{figure}

All the transverse mass distributions for Pb~+~Pb collisions for the 
different centrality classes as a function of $m_{T} - m_{0}$ 
have a very similar 
shape (see Fig.~\ref{fig:spectra1}). 
The distributions are compared more closely in 
Fig.~\ref{fig:ratioeff} where the spectra for given 
centrality classes are divided by the minimum bias spectrum. 
Still in these ratios no drastic variations are seen when displayed in 
logarithmic scale. The spectral shapes, especially for the 
intermediate centrality classes, are very similar. 
The spectra are broadened when going from peripheral 
reactions to semi-peripheral reactions, 
consistent with the nuclear enhancement seen in p~+~A   
\cite{cronin} and in S~+~Au collisions \cite{wa80:pi0:98}. However, 
the enhancement does not continue to grow for central 
collisions. Instead there is an indication of a stronger fall off in the most 
central class.

This observation is surprising, because both initial state (parton 
multiple scattering) and final state (hadron rescattering) mechanisms 
are expected to yield a further broadening of the spectra with 
decreasing impact parameter, i.e. increasing thickness or volume. 

The change in the spectrum for the most stringent centrality 
selection of $1 \% $ of the minimum bias cross section appears to be 
relatively strong, and the question arises whether this cut is really 
significant enough and whether this 
sample should not behave similarly to the adjacent sample. We have 
therefore performed detailed investigations of possible backgrounds and 
biases with respect to this particular sample.
Pile-up of multiple beam interactions is effectively suppressed by 
strong time and amplitude cuts on the trigger level. In addition, the 
total calorimeter coverage of MIRAC and the ZDC is sufficient to reject all 
possible pile-up events, as the measured energy would exceed the 
beam energy.

Also, simulations of the centrality selections including realtistic 
fluctuations in the detectors show that e.g. classes 7 and 8 are 
significantly different. This can e.g. be expressed by the mean (rms) 
of the distribution of the number of participating nucleons which are 
346 (20) and 380 (10), respectively.

As the centrality selection is performed with the transverse energy 
measured slightly off midrapidity, one might suspect that midrapidity 
particle production could suffer a different bias, if e.g. the 
pseudorapidity distribution would change significantly over the 
small range relevant here. This is, however, 
not observed. The charged particle pseudorapidity density at $\eta = 0$ 
increases in a fashion identical to the transverse energy as can be 
seen in Ref.~\cite{wa98:scaling}.

While this leads us to expect no additional bias from the event 
selection, the direct comparison of the spectra obtained with 
different analysis cuts for the most central sample (class 8) shows 
slightly larger variations than the other samples. We have therefore 
assigned a larger systematic error of $20 \% $ to this sample, as was 
discussed in Section~\ref{sec:rec}.

\begin{table}[htb]
\centering
\fbox{\begin{tabular}{|c||r@{\,}c@{}r|r|r|r|}
        \hline
        class & \multicolumn{3}{c|}{$E_{T}$ (GeV)} &
        $\sigma / \sigma_{\mathrm{mb}}$ 
        & $\left\langle N_{part} \right\rangle$ 
        & $\left\langle N_{coll} \right\rangle$ \\
        \hline \hline
        1 &  &$\le$&24.35 & 17.2~\% & $12 \pm 2 $ & $9.9 \pm 2.5$ \\
        \hline
        2 & 24.35&$-$&55.45 & 15.8~\% & $30 \pm 2$ & $30 \pm 5$ \\
        \hline
        3 & 55.45&$-$&114.85 & 18.2~\% & $63 \pm 2$ & $78 \pm 12$ \\
        \hline
        4 & 114.85&$-$&237.35 & 23.5~\% & $132 \pm 3$ & $207 \pm 21$ \\
        \hline
        5 & 237.35&$-$&326.05 & 12.3~\% & $224 \pm 1$ & $408 \pm 41$ \\
        \hline
        6 & 326.05&$-$&380.35 &  6.2~\% & $290 \pm 2$ & $569 \pm 57$ \\
        \hline
        7 & 380.35&$-$&443.20 &  5.8~\% & $346 \pm 1$ & $712 \pm 71$ \\
        \hline
        8 &  &  $>$ & 443.20 &  1.0~\% & $380 \pm 1$ & $807 \pm 81$ \\
        \hline
        6 - 8 &  &  $>$ & 326.05 &  12.7~\% & $323 \pm 1$ & $651 \pm 65$ \\
        \hline
        \end{tabular}}
        \caption{Centrality classes, as selected by the amount of 
        transverse energy measured in MIRAC, for Pb~+~Pb collisions. 
        The average number of participants and binary nucleon-nucleon
        collisions as calculated with VENUS 4.12 with an estimate of 
        the systematic error. Please note that the systematic error in 
        the number of collisions is correlated for all samples.}
        \label{tab:classes}
\end{table}

\subsection{Average $p_{T}$ and Inverse Slopes}

\begin{figure}[bt]
        \centerline{\includegraphics{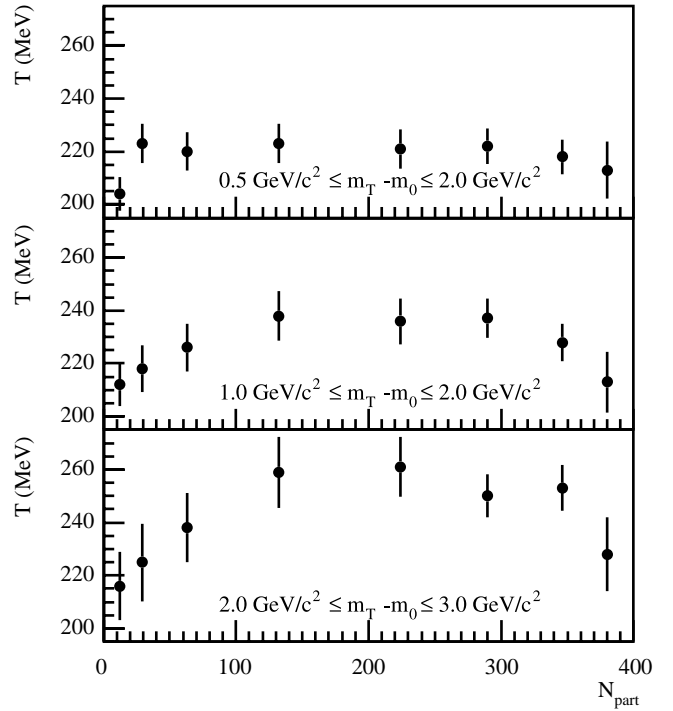}}
        \caption{Inverse slope parameters $T$ of neutral pion transverse 
        mass distributions as a 
        function of the number of participants in Pb~+~Pb collisions. The 
        exponential fits are performed in different regions $m_{T} - m_{0}$ 
        as indicated. The error bars shown contain statistical and 
        systematic errors added in quadrature.}
        \protect\label{fig:slopes}
\end{figure}

Another means to characterize the 
spectral shape is via the average transverse momentum $\left\langle 
p_{T} \right\rangle$, or via the inverse slopes $T$ of the spectra. 
The truncated average transverse momentum has 
already been presented in \cite{wa98:pi0:98}, where it was shown that 
the values of $\langle p_T(p_{T}^{min}) \rangle$, with a lower cutoff 
of $p_{T}^{min} = 0.4 \, \mathrm{GeV}/c^{2}$, increase from peripheral to medium 
central collisions but seem to saturate for still smaller impact 
parameters. Since the $\langle p_T(p_{T}^{min}) \rangle$ is  
dominated by the momentum region near the cutoff, this is not in 
contradiction with the dependence observed in Fig.~\ref{fig:ratioeff}. 
A similar analysis has been 
performed with inverse slope parameters which were obtained by 
fitting exponentials (Eq.~\ref{eq:exp})
to the spectra in limited regions of transverse mass. The extracted 
slope parameters as a function of the number of participants are shown 
in Fig.~\ref{fig:slopes}. Fits in the lowest $m_{T}$ interval yield 
a slope of $T = 204 \, \mathrm{MeV}$ for peripheral reactions. In the 
intermediate centrality range the slope appears to be constant at 
$T \approx 220 \, \mathrm{MeV}$. 

\begin{sloppypar}
Inverse slopes have also been extracted for other intervals of 
transverse mass (see Fig.~\ref{fig:slopes}). For all centralities the 
slopes increase with larger transverse mass, which is another 
demonstration of the curvature of the spectra. 
The centrality dependence is more 
pronounced for the higher $m_{T}$ regions ($1 - 2$~GeV/$c^{2}$ and 
$2 - 3$~GeV/$c^{2}$). There is a continuous rise in the 
inverse slope from very peripheral reactions up to reactions with 
$N_{part} \approx 130$, the highest slopes reaching 
$T \approx 240 \, \mathrm{MeV}$ and $T \approx 260 \, \mathrm{MeV}$, 
respectively. 
There is an indication of a decrease in the slope for very central 
collisions which is however not conclusive in view of the systematic 
errors.

From this analysis of the spectral shapes by looking at spectral 
ratios or inverse slopes one may conclude, that apart from the 
broadening in going from peripheral to semi-peripheral collisions, no 
striking features are observed. Apparently these analysis tools are 
not suited to extract information on possible more subtle variations.
\end{sloppypar}

\subsection{Scaling with System Size}

In addition to variations of the shape of the momentum spectra it is 
of interest to study the variations in absolute multiplicities. 
Especially at high transverse momentum one na{\"{\i}}vely expects an 
increase of the cross section proportional to the mass number of the 
nuclei and, correspondingly, an increase of the multiplicity 
proportional to the number of binary collisions due to the importance 
of hard scattering. In fact, it was already observed in pA collisions 
at beam energies of $ 200 - 400 \, \mathrm{GeV}$ \cite{cronin} that 
the increase in cross section at high transverse momenta is even 
stronger than the increase in the target mass.

\begin{figure}[bt]
        \centerline{\includegraphics{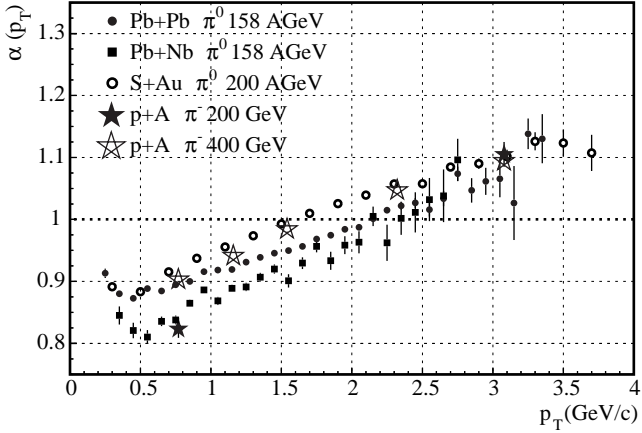}}
        \caption{Exponent $\alpha$ of the mass dependence 
        of neutral pion cross sections as a function of transverse 
        momentum. Values are obtained from ratios of minimum bias Pb~+~Pb 
        and Pb~+~Nb collisions to a parameterization of p+p collisions.
        Included for 
        comparison are results for S+Au collisions at 200$\cdot A$GeV 
        \protect\cite{wa80:pi0:98} and the proton 
        induced data from Ref.~  \protect\cite{cronin}.}
        \protect\label{fig:cronin}
\end{figure}

\begin{figure}[bt]
        \centerline{\includegraphics{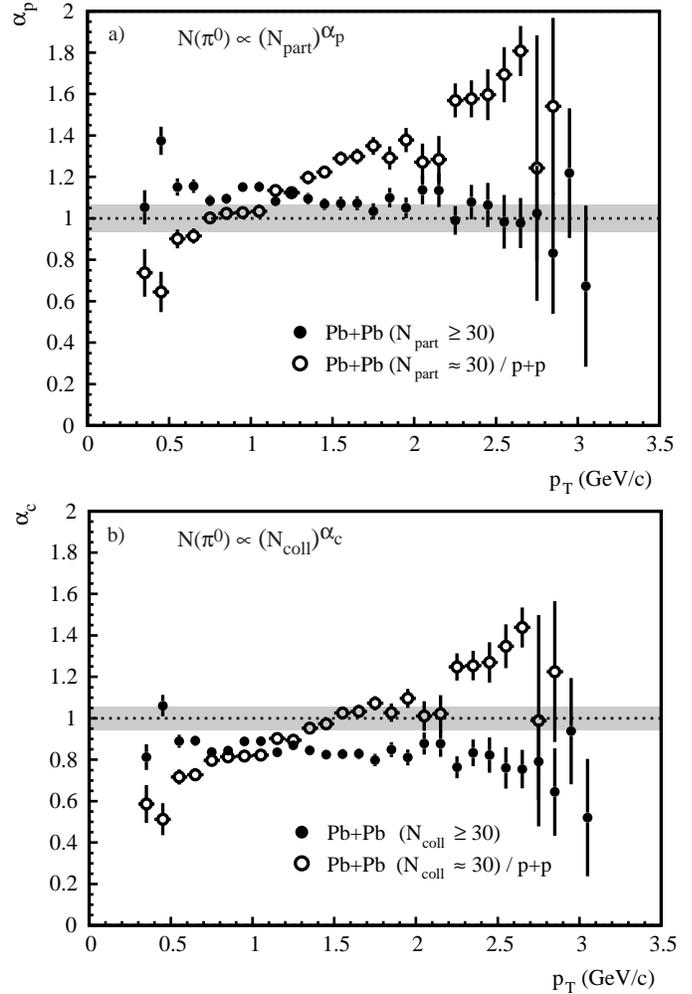}}
        \caption{Exponent $\alpha$ of neutral pion transverse 
        mass distributions parameterized as a function of the 
        number of participants (a) 
        and collisions (b) in Pb~+~Pb collisions. The 
        exponents are shown as a function of $p_{T}$. The open circles are 
        obtained from the ratio of semi-peripheral Pb+Pb data to p+p data, 
        while for the filled circles fits of a power law 
	to the seven most central classes 
        are performed. 
        The grey bands indicate 
        the systematic error relevant for this fit 
        from the calculation of the number of 
        participants and collisions and from the centrality dependent 
        systematic error of the neutral pion multiplicity. 
	The error bars are only statistical.}
        \protect\label{fig:alphanpnc}
\end{figure}

\begin{figure}[bt]
        \centerline{\includegraphics{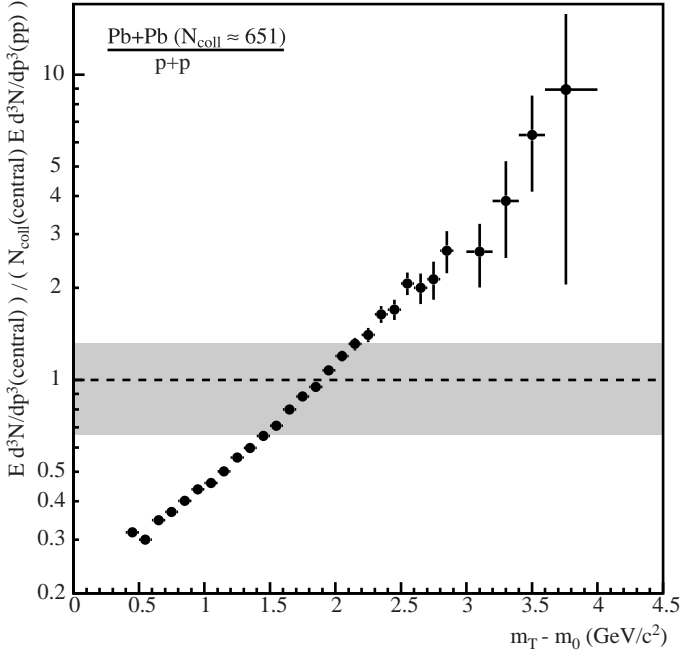}}
        \caption{Ratios of invariant multiplicity distributions of neutral 
        pions for central Pb+Pb reactions to the parameterization of 
        p+p reactions 
        normalized to the number 
        of binary collisions, also called the nuclear modification 
        factor. 
        The grey band shows the estimate of the systematic error 
	due to the calculation of the number of collisions and the absolute 
	cross section normalization relative to p+p. 
        \protect\label{fig:centpp}}
\end{figure}

This  behaviour can be investigated by parameterizing the data 
according to the phenomenological expression:
\begin{equation}
\label{eqn:cronin}
E\frac{d^3\sigma}{dp^3} ({\mathrm{A+B}}) =
(A \cdot B)^{\alpha (p_T)} \: E \frac{d^3\sigma}{dp^3} ({\mathrm{p + 
p}}),
\end{equation}
where $A$ and $B$ are the mass numbers of the projectile and target 
nuclei. For this purpose the experimental data have been compared to a 
parameterization \cite{phd:blume} of 
pp reactions \cite{don76,bus76,ang78,kou80,dem87,ada96,alp75,bre95}
scaled to the same $\sqrt{s}$:
\begin{equation}
        E \frac{d^{3}N}{dp^{3}} =  C \cdot \left( \frac{p_{0} }
	{p_{T} + p_{0} } \right)^{n},
        \label{eq:hagepp}
\end{equation}
with $C = 4.125c^{3}/ \mathrm{GeV}^{2}$, $p_{0} = 9.02 \, 
\mathrm{GeV}/c$ and $n = 55.77$. We have assigned a systematic error 
of $20 \% $ to this reference distribution.
The values of the exponents $\alpha$ for minimum bias reactions of 
Pb+Pb and Pb+Nb are shown in figure~\ref{fig:cronin}. Included for 
comparison are results for S+Au collisions at 200$\cdot A$GeV 
\cite{wa80:pi0:98} and the proton 
induced data from \cite{cronin}. All data follow the same 
trend, i.e. the exponent is considerably lower than one for low 
transverse momenta and increases monotonically with $p_{T}$ finally 
reaching values $>1$ for high $p_{T}$. However, the values differ by 
about $10 \% $ for the different systems and the point where they 
cross the line $\alpha = 1$ ranges from $p_{T} = 1.5 \, 
\mathrm{GeV}/c^{2}$ to $2.5 \, \mathrm{GeV}/c^{2}$. It should be 
noted that the different experimental trigger biases may influence 
these results.

This form of scaling analysis is only applicable to minimum bias 
reactions. To study the scaling behaviour for heavy ion 
reactions of different centralities the scaling with the number of 
participants or nucleon-nucleon collisions is more appropriate. 
An analysis of the 
scaling of the charged multiplicity with the number of participants 
has been performed in \cite{na57:scaling}. A more general analysis of 
the scaling behaviour of transverse energy and particle production in 
Pb~+~Pb collisions has been presented in \cite{wa98:scaling} and
the scaling of neutral pion production with the number of 
participants in \cite{wa98:pi0:98}. For the present paper the 
number of participants and collisions 
have been obtained as described in \cite{wa98:scaling}. 
The values are given in Table~\ref{tab:classes}. The systematic 
errors of these quantities have been obtained by comparing calculations 
with different assumptions.
In addition to VENUS and FRITIOF calculations with default parameter 
settings, we have investigated VENUS calculations with a modified 
nuclear density distribution, with a significantly worsened energy 
resolution of the calorimeter, with different assumptions on the size 
of the minimum bias cross section and ignoring the target-out 
contribution. These calculations are described in \cite{wa98:scaling}. 
Moreover we have also investigated independently results of a toy 
model calculation of nuclear geometry and also the influence of 
modest variations of the nucleon-nucleon cross section entering into the 
calculations. The maximum deviation of any of these other calculations 
from the default one is within the error given in Table~\ref{tab:classes}. 
As expected, the relative error is largest in the most peripheral 
sample. This is due to the uncertainty in modelling the 
$E_{T}$-fluctuations and the trigger threshold \textemdash effects 
which have been included in our error estimate.

It should be noted that the most peripheral data set in this analysis 
does not correspond to the most peripheral collisions theoretically 
possible from nuclear geometry, as they are rejected by the minimum 
bias trigger. Our minimum bias trigger cross section of $\approx 
6300$~mb corresponds to $\approx 85 \% $ of the expected total 
geometrical cross section. The more peripheral reactions rejected by 
our trigger would of course have a still larger uncertainty due to 
e.g. the uncertainties in the nuclear density distribution. In our 
samples the error is still at a reasonable level as quoted in the table.

Furthermore, the error in ratios of e.g. the number of collisions for 
different samples is considerably smaller, since some of the systematic 
effects are correlated, as is obvious e.g. for the size of the 
nucleon-nucleon cross section.

The scaling of the neutral pion production is analyzed both
as a function of the number of collisions and as function of
the number of participants. As 
in the earlier publications the yield is parameterized as:
\begin{equation}
        \label{eqn:scalenp}
        E\frac{d^3N}{dp^3} (N_{part}) \propto (N_{part})^{\alpha_{p}} 
\end{equation}
or
\begin{equation}
        \label{eqn:scalenc}
        E\frac{d^3N}{dp^3} (N_{coll}) \propto (N_{coll})^{\alpha_{c}}.
\end{equation}
The results are presented in Fig.~\ref{fig:alphanpnc} where the upper 
panel shows the scaling with the number of participants and the lower 
panel with the number of collisions.  The open 
circles represent the exponents calculated from the ratio of 
semi-peripheral collisions (class 2) to the parameterization of pp 
collisions (Eq.~\ref{eq:hagepp}). In this case a behaviour very similar to the 
original Cronin-effect is observed: The exponents increase 
monotonically with increasing $p_{T}$ reaching values $> 1$ for 
large $p_{T}$. A scaling of the particle yields with 
$(N_{coll})^{\alpha_{c}=1}$ 
is equivalent to a scaling of the minimum bias cross sections with 
$A^{\alpha=1}$. This value is reached in Fig.~\ref{fig:alphanpnc}b at 
$p_{T} \approx 1.5 \, \mathrm{GeV}/c^{2}$, similar to where the scaling 
exponents of the p+A and S+Au data in Fig.~\ref{fig:cronin} reach the value 
of 1. However, the minimum bias Pb+Pb data in Fig.~\ref{fig:cronin} show a 
weaker scaling exponent at intermediate and higher $p_{T}$.

\begin{figure}[bt]
        \centerline{\includegraphics{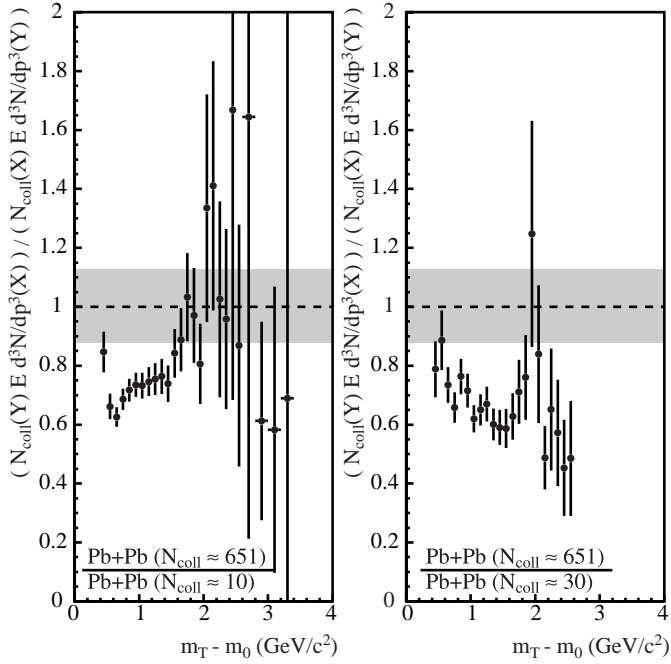}}
        \caption{Ratios of invariant multiplicity distributions of neutral 
        pions for 
        central to peripheral Pb+Pb collisions 
        normalized to the number 
        of binary collisions. The left plot shows the ratio using the 
        most peripheral sample, the right plot a similar ratio with 
        the second most peripheral sample. 
        The grey bands show the estimate of the systematic error 
	due to the calculation of the number of collisions and due to the 
	systematic error in the $\pi^{0}$ distribution. 
        \protect\label{fig:centperi}}
\end{figure}

\begin{sloppypar}
Fits to the centrality classes 2-8 (displayed 
as filled circles in Fig.~\ref{fig:alphanpnc}), i.e. up to 
the most central collisions, show almost constant exponents for all 
transverse momenta. In fact, the values even indicate a slight 
decrease with increasing $p_{T}$. At the highest $p_{T}$ the 
exponents appear to be $\alpha_{p} \approx 1$ and $\alpha_{c} \approx 
0.8$. They are considerably below the scaling exponents for 
semi-peripheral Pb+Pb relative to p+p. Noticeably the 
scaling exponent $\alpha_{c}$ is also significantly below the value 
of 1 expected as a na{\"\i}ve scaling for hard scatterings.
\end{sloppypar}

\begin{figure}[bt]
        \centerline{\includegraphics{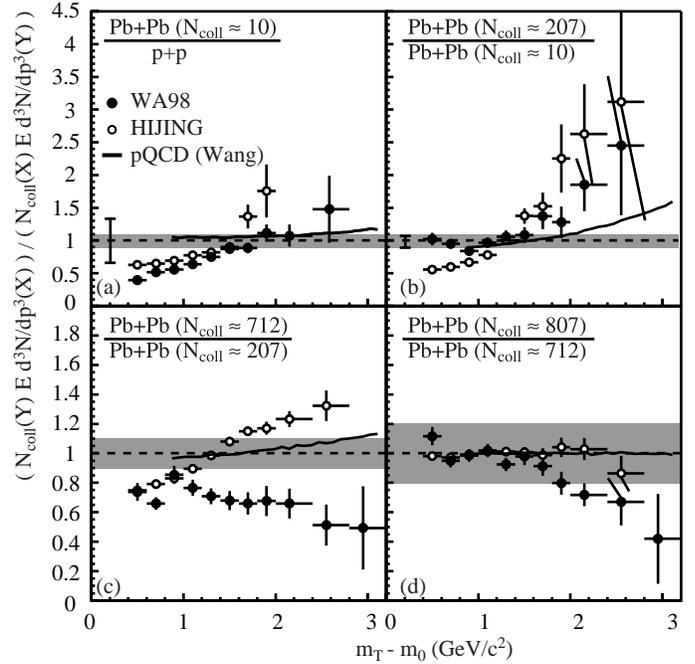}}
        \caption{Ratios of invariant multiplicity distributions of neutral pions 
        normalized to the number 
        of binary collisions. The upper left plot (a) shows the ratio of 
        peripheral Pb+Pb collisions to the parameterization of p+p 
        [\protect\ref{eq:hagepp}]. The other plots (b-d) show
        different ratios of a more 
        central sample to a more peripheral sample of Pb+Pb collisions. The 
        filled circles show the experimental results, the 
        open circles are results from the HIJING event generator  
        and the solid lines 
         similar ratios from pQCD calculations 
        including $p_{T}$ broadening \protect\cite{wang:pqcd2}. 
        The grey bands show the estimate of the systematic error 
	due to the 
	systematic error in the $\pi^{0}$ distribution. The additional error bar 
	in (a) and (b) shows the normalization uncertainty 
	due to the calculation of the number of collisions and the absolute 
	cross section normalization relative to p+p.
        \protect\label{fig:rationew}}
\end{figure}

\begin{figure}[bt]
        \centerline{\includegraphics{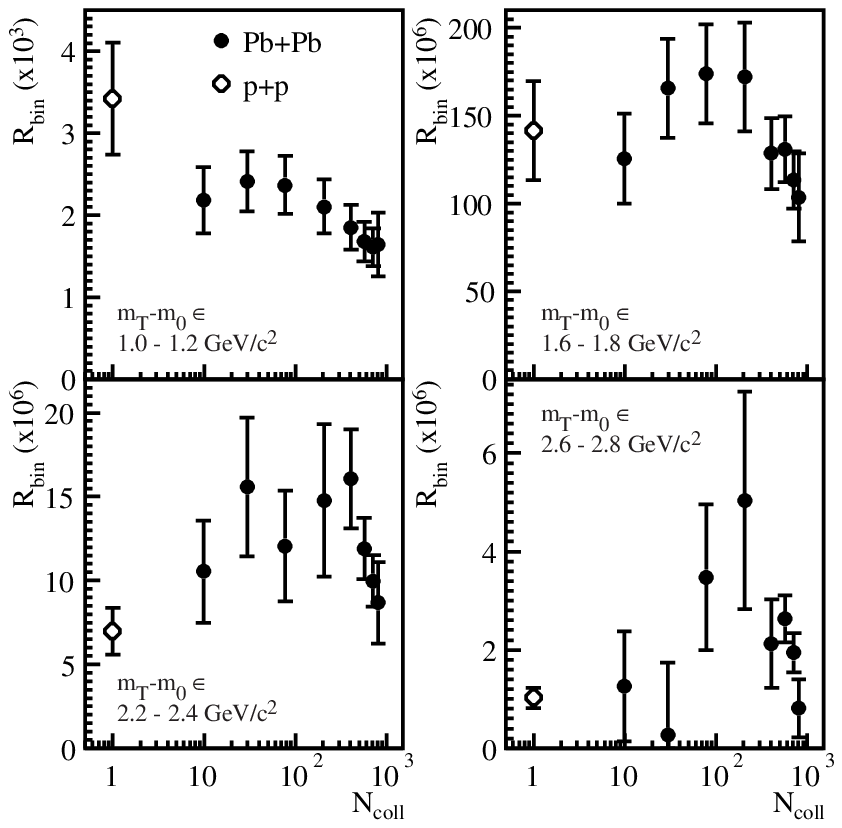}}
        \caption{Multiplicities of neutral pions 
        normalized to the number 
        of binary collisions $R_{bin} = (E d^3N/dp^3)/ N_{coll}$ 
	as a function of $N_{coll}$ 
        for different ranges of transverse mass as indicated in the 
        pictures. 
	The filled circles show the experimental results for Pb+Pb reactions, the 
        open circles the parameterization for p+p reactions 
        (Eq.~\ref{eq:hagepp}).  
        The error bars contain statistical and 
        systematic errors added in quadrature.
        \protect\label{fig:ratiopt}}
\end{figure}

Ratios of the measured pion multiplicity distributions for two 
different samples (labeled X and Y)
normalized to the number of collisions given in 
Table~\ref{tab:classes}: 
\begin{equation}
        \label{eqn:nuclmod1}
        R_{XY}(m_{T}) \equiv 
	\frac{{\left( E\frac{d^3N}{dp^3} (m_{T}) / N_{coll} \right)} _{X}}
	{{\left( E\frac{d^3N}{dp^3} (m_{T}) / N_{coll} \right) }_{Y}}
\end{equation}
are shown in Figs.~\ref{fig:centpp}, \ref{fig:centperi} and \ref{fig:rationew}. 
In Fig.~\ref{fig:centpp} the ratio of the $12.7 \% $ most central 
collisions to the parameterization of p+p is shown:
\begin{equation}
        \label{eqn:nuclmod2}
        R_{X}(m_{T}) \equiv 
	\frac{{\left( E\frac{d^3N}{dp^3} (m_{T}) / N_{coll} \right)} _{X}}
	{{\left( E\frac{d^3N}{dp^3} (m_{T}) \right) } _{pp}}.
\end{equation}
This special case of equation \ref{eqn:nuclmod1} 
is sometimes referred to as the \emph{nuclear modification 
factor} \cite{wang:pqcd2}. The ratio is $ \approx 0.3$ at low $m_{T}$ 
and increases exponentially towards higher $m_{T}$ approaching values 
close to 10. \footnote{The same ratio for central Pb+Pb(Au) collisions 
is shown in \protect\cite{wang:pqcd2}. There a curve is drawn which 
saturates at $R=2$ for high $p_{T}$. It is actually dominated by data 
points at $p_{T} = 2 - 3 \,\mathrm{GeV}/c$, and the WA98 data points 
can be seen to lie above the curve at high $p_{T}$. Remaining 
differences in the two figures might be due to different estimates for 
the pp-spectra being used in the ratios.}
This is in line with the earlier observations and demonstrates 
again that relative to p+p reactions there is a strong Cronin effect in heavy 
ion collisions for all centralities. 
However, as evident from Fig.~\ref{fig:alphanpnc}b, most of the 
anomalous enhancement is already present in peripheral reactions, and 
the additional enhanement in central reactions is much weaker. In the 
following we will concentrate on this particular observation.

Fig.~\ref{fig:centperi} shows ratios of the $12.7 \% $ most central 
collisions to the two most peripheral samples. Both ratios are below 1 
for low $m_{T}$. 
In comparison to the most peripheral sample (left 
part) the spectra are compatible with a scaling with the number of 
collisions in central reactions at higher $m_{T}$, while relative to 
the semi-peripheral class (right part) 
the enhancement is significantly weaker even 
at high $m_{T}$. 
Similar ratios for more detailed centrality 
selections are shown in Fig.~\ref{fig:rationew}.
The ratio of peripheral 
Pb+Pb collisions to p+p (Eq.~\ref{eq:hagepp})
(Fig.~\ref{fig:rationew}a) increases 
strongly with increasing transverse mass -- this is in line with the 
Cronin effect discussed above. A similar trend is observed when going 
from peripheral to medium-central data (Fig.~\ref{fig:rationew}b). 
In addition, the pion 
production is seen to increase roughly proportional to the number of collisions 
even at low transverse masses. Going from medium central to central 
(Fig.~\ref{fig:rationew}c) 
the trend is reversed: the ratio decreases with increasing transverse 
mass as was already seen in Fig.~\ref{fig:ratioeff} and the 
pion multiplicities increase more weakly than the number of collisions. 
The ratio of very central to central collisions shows an indication of 
a similar effect although not very significant.

Included in Fig.~\ref{fig:rationew} are results of HIJING 
calculations (open circles) for the same centralities. They show a very 
different trend: For all but the most central case the ratios 
increase with increasing transverse mass, for high transverse masses 
the ratio is always larger than one.  
HIJING does not describe p+p data well at these energies while a reasonable 
description of central Pb+Pb collisions is obtained by an 
implementation of the Cronin effect via a soft transverse momentum 
kick model which introduces a very strong $A$ dependence, as was pointed 
out already in \cite{gyulassy98}. Thus 
the reasonable description of central Pb+Pb 
collisions by HIJING appears to be fortuitous. The authors of 
\cite{gyulassy98} stated that a better implementation of the 
Cronin effect should use an additional intrinsic $p_{T}$ of the 
incoming partons and a likely weaker $p_{T}$ broadening. Still, in 
any model of the Cronin effect one would expect the ratios of central
to less central spectra, as shown in 
Fig.~\ref{fig:rationew}c, to be larger than one at high transverse 
mass. Similarly, one would expect the scaling exponent  
shown in Fig.~\ref{fig:alphanpnc} to be $\alpha_{c}>1$ at high 
transverse momentum.
A more refined calculation including intrinsic $p_{T}$ was performed 
in \cite{wang:pqcd2} using the same model as in \cite{wang:1998:qcd}. 
The results are also shown in Fig.~\ref{fig:rationew} as solid lines. 
As expected, the ratios show a weaker increase at high $m_{T}$ 
compared to HIJING, but are all $\ge 1$ and thus do not explain the 
centrality dependence as seen in Fig.~\ref{fig:rationew}c.

With these observations in mind one can revisit the evolution of the 
particle yields with system size at given transverse mass. This is 
done in Fig.~\ref{fig:ratiopt} where the neutral pion yield per binary 
collision:
\begin{equation}
        \label{eqn:yieldbin}
        R_{bin}\equiv E\frac{d^3N}{dp^3}/ N_{coll}
\end{equation}
is shown as a function of the number of collisions for four 
different transverse mass intervals. In addition to Pb+Pb reactions 
also the parameterization of p+p is included. 
At relatively low transverse mass the yield per collision decreases as 
expected from the scaling exponent $\alpha_{c}<1$. With increasing  
transverse mass a rise of $R_{bin}$ develops which is most prominent 
for the highest $m_{T}$. However, it is also clear that $R_{bin}$ 
decreases significantly with $N_{coll}$ for $N_{coll} > 200$. The 
exact point of
turnover of the evolution of the pion yields could not be so easily 
observed in the investigation of the scaling exponents above -- it is much 
more clearly born out when normalizing to the number of binary 
collisions as in $R_{bin}$.

\section{Summary}

We have presented transverse mass spectra of neutral pions 
in Pb+Pb reactions at 
158$\cdot A$~GeV for different centralities. A comparison to several event 
generators found none of them able to adequately describe the 
spectra. 

While there is a strong broadening of the transverse mass spectra
in going from p+p to peripheral Pb+Pb reactions, and further to  
semi-peripheral reactions, there is an indication of a stronger fall 
off of the spectra when going to the most central selections. The 
absolute yields at high transverse momenta show an enhancement 
which grows stronger than with the number of collisions up to medium 
central Pb+Pb 
collisions, which is qualitatively similar to the Cronin effect 
observed in p+A collisions. For central collisions, however, the 
further increase in multiplicity at high $p_{T}$ is weaker than with the 
number of collisions. This is in qualitative contradiction to 
conventional explanations of the Cronin effect ($p_{T}$-broadening) 
which is expected to 
cause a further strengthening of the nuclear enhancement with more 
central collisions. 
As a possible explanation of this behaviour one may consider that the 
multiple scattering 
mechanisms which are expected to be responsible for the 
apparent $p_{T}$-broadening in pA and peripheral Pb+Pb might be modified 
in central Pb+Pb collisions. This might be possible if both \emph{initial} 
and \emph{final} state scatterings are relevant for the enhancement, 
and the relative contributions are shifted more towards final state 
contributions in central collisions, reminiscent of a more and more 
thermalized system. 
Alternative explanations might involve suppression 
mechanisms independent of the nuclear enhancement in question, e.g. 
an onset of quenching via energy loss of produced 
particles (partons or hadrons) in central Pb+Pb collisions. 

\begin{acknowledgement}
We wish to express our gratitude to the CERN accelerator division for
excellent performance of the SPS accelerator complex. We acknowledge with
appreciation the effort of all engineers, technicians and support staff 
who have participated in the construction of the experiment.

\begin{sloppypar}
This work was supported jointly by 
the German BMBF and DFG, 
the U.S. DOE,
the Swedish NFR and FRN, 
the Dutch Stichting FOM, 
the Polish KBN under  Contract No 621/E-78/SPUB-M/CERN/P-03/DZ211/,
the Grant Agency of the Czech Republic under contract No. 202/95/0217,
the Department of Atomic Energy,
the Department of Science and Technology,
the Council of Scientific and Industrial Research and 
the University Grants 
Commission of the Government of India, 
the Indo-FRG Exchange Program,
the PPE division of CERN, 
the Swiss National Fund, 
the INTAS under Contract INTAS-97-0158, 
ORISE, 
Research-in-Aid for Scientific Research
(Specially Promoted Research \& International Scientific Research)
of the Ministry of Education, Science and Culture, 
the University of Tsukuba Special Research Projects, and
the JSPS Research Fellowships for Young Scientists.
ORNL is managed by Lockheed Martin Energy Research Corporation under
contract DE-AC05-96OR22464 with the U.S. Department of Energy.
The MIT group has been supported by the US Dept. of Energy under the
cooperative agreement DE-FC02-94ER40818.
\end{sloppypar}
\end{acknowledgement}

\end{document}